# Role of covalent hybridization in martensitic structure and magnetic properties of shape memory alloys: the case of $Ni_{50}Mn_{5+x}Ga_{35-x}Cu_{10}$

G. J. Li,[1] E. K. Liu,[1] H. G. Zhang,[1] Y. J. Zhang,[1] G. Z Xu,[1] H. Z. Luo,[2] H. W. Zhang,[1] W. H. Wang,[1,a)] and G. H. Wu[1]

*1 Beijing National Laboratory for Condensed Matter Physics, Institute of Physics, Chinese Academy of Sciences, Beijing 100190, People's Republic of China*

*2 School of Material Science and Engineering, Hebei University of Technology, Tianjin 300130, People's Republic of China*

Abstract：

We have investigated the impact of covalent hybridization on martensitic structure and magnetic properties of $Ni_{50}Mn_{5+x}Ga_{35-x}Cu_{10}$ shape memory alloys. We found that the lattice distortion ((c-a)/a) of $L1_0$ martensite monotonously changes with the substitution of Mn for Ga atoms and shows a kink behavior at Ga(at.%)= 25 due to the weakened covalent effect between main-group and transition-metal atoms. Moreover, owing to the competition between covalence hybridization and magnetic ordering of introduced Mn atoms, the molecular magnetic moment and Curie temperature coincidently show maximums at Ga(at.%)=25 as well. These behaviors are closely associated with corresponding changes of the strength of covalent hybridization. The results therefore suggest that careful control of the concentration of main-group atoms in Heusler alloys can serve as an additional general tuning parameter for searching new multifunctional materials.

---

a) Author to whom correspondence should be addressed; electronic mail: wenhong.wang@iphy.ac.cn

Heusler alloys are ternary compounds with the stoichiometric formula $X_2YZ$, where X and Y are transition metals and Z is main-group metal, respectively. In these alloys, the *p-d* orbital hybridization between main-group and transition-metal atoms determines phase stability[1-5], magnetic properties[6-8], lattice parameter[6, 9] and atomic occupations[10]. For ferromagnetic shape memory Heusler alloys, the intensity of *p-d* orbital hybridization largely depends on the number of main-group atoms, which can significantly influence the electronic structure at the Fermi level and thus the phase stability. Previously Zayak [1, 2], Roy [3] and Bhobe [4, 5] found that in the $Ni_2MnZ(Z=Ga,In,Sn)$ and Cu doped $Ni_2MnGa$ ferromagnetic shape Heusler alloys, the *p-d* orbital hybridization can change the distribution of electrons near Fermi level, affecting the density of state near Fermi level and modifying the phase stability[11]. However, it is well known that the martensitic structure is sensitive to the composition. With the changes of composition, the well-studied NiMnGa alloys can show five-layer modulated (5M), seven-layer modulated (7M) or non-modulated ($L1_0$) martensitic structure at room temperature[12, 13]. Owing to the complexness and variations of the martensitic structure, it is not convenient to study the influence of covalent effect on the change of martensitic structure and magnetic properties. Furthermore, in martensite, the distorted lattice can strongly modifies the covalent hybridization between transition metal and main group atoms[11, 14]. So far, however, there are few reports regarding this effect on the crystal structure and magnetic properties of martensite.

Here, we choose $Ni_{50}Mn_{5+x}Ga_{35-x}Cu_{10}$ alloys as a representative case to study the effect of covalent hybridization on the martensitic structure and magnetic properties. Previously, several physical properties in $Ni_2Mn_{1-x}Cu_xGa$ alloys have been investigated from first principles calculations and experiments as well[15-18]. By addition Cu in $Ni_2Mn_{1-x}Cu_xGa$ alloys, Roy et al.[3] have suggested an increased covalence of Ni atom and therefore the enhance of covalence hybridization between Ni and Ga leads martensitic transformation temperature to a high level. Therefore, the Ni-Mn-Ga-Cu alloys with high Cu-content provide a convenient platform to study the martensitic structure at room temperature.

Polycrystalline ingots of $Ni_{50}Mn_{5+x}Ga_{35-x}Cu_{10}$ ($x=0\sim15$) alloys were prepared by arc-melting high-purity metal Ni,Mn,Ga and Cu under argon atmosphere. All the ingots were annealed in evacuated quartz tubes with argon at 1073 K for three days and then quenched into ice water. The X-ray diffraction (XRD) with Cu-Kα was employed to characterize the crystal structure and to determine the lattice parameter. The martensitic transformation temperatures were determined by DSC. The saturation magnetizations and Curie temperatures of all samples were measured by a superconducting quantum interference device (SQUID) magnetometer. The Korringa-Kohn-Rostoker coherent-potential approximation and local density approximation (KKR-CPA-LDA) method[19-21] was applied to calculate the magnetic structures of $Ni_{50}Mn_{5+x}Ga_{35-x}Cu_{10}$ ($x=0\sim15$) alloys. The ideal lattice parameters for first-principles calculations were obtained by linearly fitting the experimental results. The valence-electron localization function was calculated by CASTEP package[22, 23].

In order to investigate the covalent hybridization in Heusler alloys, the first thing is to analyze the electronic distribution. Here we prefer to simplify the landscape by plotting the valence-electron localization function contour map in the (100) plane of Ni$_2$MnGa alloy in martensitic state, displayed in Fig. 1 (b). There is strong *p-d* covalent hybridization between the nearest neighbor Ni and Ga atoms, similar to the case in MnNiGe:Fe system[23]. This hybridization is believed to dominate martensitic transformation temperature in both NiMnGa and NiMnGaCu alloys, which was proved both experimentally and theoretically[3, 14, 24]. And the *p-d* hybridization effect on the lattice parameters was revealed in our recent results[6, 23]. Based on these results, once the hybridization is established, the decrease of Ga content will weaken this hybridization effect and consequently result in the changes of martensitic transformation temperature, lattice parameter and atomic occupation. In this work, the substitution of Mn for Ga atoms in Ni$_{50}$Mn$_{5+x}$Ga$_{35-x}$Cu$_{10}$ alloys raises the valence electrons concentration, which will result in convergence of electrons near Fermi level and enhance martensitic transformation temperature higher than room temperature. Figure 2 shows the martensitic lattice parameters of Ni$_{50}$Mn$_{5+x}$Ga$_{35-x}$Cu$_{10}$ alloys as a function of Ga content. All lattice parameters were obtained by indexing the XRD patterns (see supplementary Fig.S1). With the decrease of Ga content in Ni$_{50}$Mn$_{5+x}$Ga$_{35-x}$Cu$_{10}$ alloys, the lattice parameter a (=b) decreases while c increases monotonously, which results in the increase of lattice distortion ((c-a)/a). Linearly fitting these experimental data, we find that the slopes of the lattice parameter a (=b) and lattice distortion both suddenly change at Ga=25(at.%), showing kink behaviors, as marked by the arrow in Fig. 2. This kink behavior of unit-cell size can be ascribed to the weakened covalent hybridization between transition-metal and main-group atoms due to the decrease of Ga content [6].

Different from the $L1_0$ martensitic structure in $Ni_{50}Mn_{17.5}Ga_{22.5}Cu_{10}$ alloy, the Cu-free $Ni_{50}Mn_{27.5}Ga_{22.5}$ alloy possesses the 5M martensitic structure[12]. This 5M structure can result from the substitution of smaller Cu atom (0.157nm) for larger radius Mn atom (0.179nm), which causes other atoms around newly introduced Cu slightly shift and consequently destroy the periodic arrangement of the atoms[25]. As a result, the hybridization between Ni and Ga was reinforced and results in the increase of martensitic transformation temperature[3, 14]. This is the physical mechanism why the martensitic transformation temperature of $Ni_{50}Mn_{17.5}Ga_{22.5}Cu_{10}$ is higher than that of $Ni_{50}Mn_{27.5}Ga_{22.5}$ alloy.

Figure 3 (a) shows Ga content dependence of molecular magnetic moment of $Ni_{50}Mn_{5+x}Ga_{35-x}Cu_{10}$ alloys. We found that the molecular magnetic moment firstly increases with the decreasing of Ga content, reaching a maximum value at Ga (at.%) = 25, and then decreases with the further substitution of Mn for Ga atoms. In order to understand the anomalous variation of molecular magnetic moment, we should first determine the atomic occupation during the substitution. In Heusler alloys, the covalent hybridization between transition-metal and main-group atoms determines the atomic occupations. Generally, the main-group atoms mainly occupy *D* sites, the transition-metal atoms with relatively more valence electrons will preferentially occupy the *A* and (or) *C* sites while those with relatively less valence electrons occupy the *B* sites[6, 10, 26]. In $Ni_{50}Mn_{5+x}Ga_{35-x}Cu_{10}$ alloys, the valence-electron number of transition metal Cu ($3d^{10}4s^1$) atom is larger than those of Ni($3d^84s^2$), Mn($3d^54s^2$) and Ga($4s^2p^1$) atoms. The details about atomic preferential occupation and arrangement type of $Ni_{50}Mn_{5+x}Ga_{35-x}Cu_{10}$ alloys are as follows and clearly shown in Fig.3 (b): (i)

When Ga(at.%)=25, Ga atoms just fully occupy the *D* sites. The Cu and partial Ni atoms occupy the *A* and *C* sites. As a result, the rest of Ni atoms and the Mn atoms will occupy the *B* sites; (ii) When Ga (at.%) > 25, the rich Ga atoms will occupy the *B* sites except that most of them occupy the *D* sites. (iii) When Ga (at.%) < 25, the newly introduced Mn atoms will occupy D site theoretically. Here we should point out that the atomic occupations in tetragonal martensite structure are the same as those in cubic austenite structure due to the diffusionless martensitic transformation.

From the atomic occupation and magnetic properties mentioned above, we find that, in $Ni_{50}Mn_{5+x}Ga_{35-x}Cu_{10}$ alloys, the magnetic properties depend on the competition and balance of the magnetic ordering of Mn atoms and covalent hybridization between Ga and transition metals. When Ga(at.%)>25, the stronger covalent hybridization drives all the Mn atoms to occupy the third nearest neighboring sites of each other and to spontaneously parallelly align. So, in this strong covalent hybridization circumstance, the molecular magnetic moment of $Ni_{50}Mn_{5+x}Ga_{35+x}Cu_{10}$ alloys increases with the increase of Mn content. When Ga(at.%)<25 in $Ni_{50}Mn_{5+x}Ga_{35+x}Cu_{10}$ alloys, the moments of newly introduced Mn atoms prefer to occupy *D* sites (next nearest neighbor of *B* sites) and antiparallelly align to original Mn moments due to the weakened *p-d* covalent hybridization, which results in the unusual decrease in molecular magnetic moment in $Ni_{50}Mn_{5+x}Ga_{35+x}Cu_{10}$ alloys.

Based on the atomic preferential occupation rule mentioned above, first-principles calculations have been carried out by KKR-CPA-LDA method[19-21] to simulate the magnetic structure and the calculated data are plotted in Fig. 3(a). We found that the calculated molecular magnetic moments are well consistent with the experimental values. In Table I, we listed the atomic magnetic moments with various

Ga contents. One can find that: (i) the atomic moments of Mn(*B*) and Mn(*D*) are larger than those of Ni(*A,C*) and Ni(*B*)(<1$\mu_B$), which are more than 3.2 $\mu_B$. Thus, they are the main contributor of the magnetism in $Ni_{50}Mn_{5+x}Ga_{35-x}Cu_{10}$ alloys. (ii) Atomic magnetic moments of Ni and Mn (*B*) are parallelly aligned to each other when Ga (at.%) > 25, while atomic moments of Ni and Mn(*B*) atoms are anti-parallelly aligned to those of newly introduced Mn (*D*) atoms when Ga (at.%) < 25. (iii) Atomic magnetic moments of Ni (*A,C*) show the same variation tendency as the molecular magnetic moment in $Ni_{50}Mn_{5+x}Ga_{35-x}Cu_{10}$ alloys, simultaneously reaching a maximum when Ga (at.%) =25. Due to the nearest neighboring relationship between the atoms in *A*/*C* sites and *B*/*D* sites, as shown in the inset of Fig.3 (a), the magnetic moment of the atoms in *A* and/or *C* sites will be the most sensitive to the changeable covalent effect. Thus, in $Ni_{50}Mn_{5+x}Ga_{35-x}Cu_{10}$ alloys, the change behavior of atomic moments of Ni (*A,C*) can be ascribed to the changes of covalence hybridization between Ga and transition-metal atoms.

Figure 4 shows the structural and magnetic phase diagram of $Ni_{50}Mn_{5+x}Ga_{35-x}Cu_{10}$ alloys. One can find that the $T_C^M$ firstly increases from 24 K (for Ga(at.%)=35 alloy) with decreasing of Ga content, showing a local maximum value of 269 K at Ga(at.%)=25, and then turns to decreasing with the further substitution of Ga for Mn atoms. This change behavior of $T_C^M$ can reveal the variation of exchange interaction in $Ni_{50}Mn_{5+x}Ga_{35-x}Cu_{10}$ alloys. When Ga (at.%)>25 in $Ni_{50}Mn_{5+x}Ga_{35-x}Cu_{10}$ alloys, according to the atomic preferential occupation and magnetic ordering mentioned above, the number of magnetic atoms in ferromagnetic sublattice increases with the substitution of Mn for Ga atoms. Thus, the ferromagnetic exchange interaction between magnetic atoms is established gradually and becomes stronger, which results in the increased $T_C^M$. However, when Ga(at.%)<25, the anti-ferromagnetic sublattice of Mn(*D*) occurs in the original ferromagnetic matrix, which destroys the original ferromagnetic sublattice and leads to the decreases of subsequent $T_C^M$. The similar results have been reported in $Mn_2CoNi_xGa_{1-x}$ alloys, due to the formation of anti-ferromagnetic sublattice in the ferromagnetic matrix[26].

In summary, we have used a combination of experimental and theoretical techniques to address the importance of covalent hybridization on the martensitic structure and magnetic properties of $Ni_{50}Mn_{5+x}Ga_{35-x}Cu_{10}$(*x*=0-15) alloys. We have shown that, as a result of *p-d* orbital hybridization between Ga and transition-metal atoms, the lattice parameter(a=b) and lattice distortion((c-a)/a) of $Ni_{50}Mn_{5+x}Ga_{35-x}Cu_{10}$ alloys monotonously changes with the decrease of Ga content and coincidently shows kink behaviors at Ga(at.%)=25. This kink behavior can be attributed to the weakened covalence effect due to the decrease of Ga atoms. Moreover, the competition between covalent hybridization and magnetic ordering of newly introduced Mn atoms makes the molecular magnetic moment and Curie temperature of these alloys coincidently shows Λ-shaped changes behavior, reaching maximums at Ga(at.%) =25. The higher martensitic transformation temperatures and wider temperature hysteresis were observed in Ga poor alloys, making the $Ni_{50}Mn_{5+x}Ga_{35-x}Cu_{10}$($x\cong 15$) alloys as the single–phase wide-hysteresis high-temperature shape memory alloys.

This work is supported by the National Natural Science Foundation of China in Grant No. 51021061, 51071172 and 11174352 and National Basic Research Program of China (973 Program, 2012CB619405).

References:

[1] A. T. Zayak, P. Entel, K. M. Rabe, W. A. Adeagbo, and M. Acet, Phys. Rev. B **72**, 054113 (2005).
[2] A. T. Zayak, W. A. Adeagbo, P. Entel, and K. M. Rabe, Appl. Phys. Lett. **88**, 111903 (2006).
[3] S. Roy, E. Blackburn, S. M. Valvidares, M. R. Fitzsimmons, S. C. Vogel, M. Khan, I. Dubenko, S. Stadler, N. Ali, S. K. Sinha, and J. B. Kortright, Phys. Rev. B **79**, 235127 (2009).
[4] P. A. Bhobe, K. R. Priolkar, and P. R. Sarode, J. Phys.: Condens. Matter **20**, 015219 (2008).
[5] P. A. Bhobe, K. R. Priolkar, and P. R. Sarode, J. Phys. D: Appl. Phys. **41**, 045004 (2008).
[6] G. J. Li, E. K. Liu, H. G. Zhang, J. F. Qian, H. W. Zhang, J. L. Chen, W. H. Wang, and G. H. Wu, Appl. Phys. Lett. **101**, 102402 (2012).
[7] H. M. Hong, Y. J. Kang, J. Kang, E. C. Lee, Y. H. Kim, and K. J. Chang, Phys. Rev. B **72**, 144408 (2005).
[8] S. E. Kulkova, S. V. Eremeev, and S. S. Kulkov, Solid State Commun. **130**, 793 (2004).
[9] B. Joachim, B. Benjamin, H. F. Gerhard, S. Hryhoriy, G. Andrei, N. Shahab, and F. Claudia, J. Phys. D:Appl. Phys. **42**, 185401 (2009).
[10] T. J. Burch, T. Litrenta, and J. I. Budnick, Phys.Rev.lett. **33**, 421 (1974).
[11] P. A. Bhobe, K. R. Priolkar, and P. R. Sarode, Phys. Rev. B **74**, 224425 (2006).
[12] C. Jiang, Y. Muhammad, L. Deng, W. Wu, and H. Xu, Acta Mater. **52**, 2779 (2004).
[13] R. K. Singh, M. Shamsuddin, R. Gopalan, R. P. Mathur, and V. Chandrasekaran, Mater. Sci. Eng. A **476**, 195 (2008).
[14] K. R. Priolkar, P. A. Bhobe, and P. R. Sarode, Adv. Mater. Res. **52**, 155 (2008).
[15] S. Stadler, M. Khan, J. Mitchell, N. Ali, A. M. Gomes, I. Dubenko, A. Y. Takeuchi, and A. P. Guimaraes, Appl. Phys. Lett. **88** (2006).
[16] J. F. Duan, Y. Long, B. Bao, H. Zhang, R. C. Ye, Y. Q. Chang, F. R. Wan, and G. H. Wu, J. Appl. Phys. **103** (2008).
[17] M. Khan, I. Dubenko, S. Stadler, and N. Ali, J. Appl. Phys. **97**, 10M304 (2005).
[18] J. M. Wang, H. Y. Bai, C. B. Jiang, Y. Li, and H. Xu, Mater. Sci. Eng. A **527**, 1975 (2010).
[19] S. Blügel, H. Akai, R. Zeller, and P. H. Dederichs, Phys. Rev. B **35**, 3271 (1987).
[20] S. Kaprzyk and A. Bansil, Phys. Rev. B **42**, 7358 (1990).
[21] H. Akai, Phys. Rev. Lett. **81**, 3002 (1998).
[22] S. J. Clark, M. D. Segall, C. J. Pickard, P. J. Hasnip, M. I. J. Probert, K. Refson, and M. C. Payne, Z. Kristallogr. **220**, 567 (2005).
[23] E. K. Liu, W. H. Wang, L. Feng, W. Zhu, G. J. Li, J. L. Chen, H. W. Zhang, G. H. Wu, C. B. Jiang, H. B. Xu, and F. de Boer, Nat. Commun. **3**, 873 (2012).
[24] S. Özdemir Kart, M. Uludoğan, I. Karaman, and T. Çağın, Phys.Stat.Sol.(a) **205**, 1026 (2008).
[25] J. M. Wang and C. B. Jiang, Scripta Mater. **62**, 298 (2010).
[26] L. Ma, W. H. Wang, C. M. Zhen, D. L. Hou, X. D. Tang, E. K. Liu, and G. H. Wu, Phys. Rev. B **84**, 224404 (2011).

Figure Captions:

FIG. 1. (Color online) (a) Crystal structure of $Ni_2MnGa$ Heusler alloy in the $L1_0$ martensite state. (b) Valence-electron localization function contour map in the (100) plane. The scale bar on the bottom right corner shows the increasing covalence bonding when the color changes from blue to red.

FIG. 2. (Color online) The composition dependence of lattice parameter and lattice distortion (c-a/a) of the alloys, where the straight lines are from linearly fitting experimental values.

FIG. 3. (Color online) (a) Composition dependence of experimental (solid triangle) and calculated (open triangle) molecular magnetic moment in $Ni_{50}Mn_{5+x}Ga_{35-x}Cu_{10}$ alloys. The inset shows the nearest neighbor atomic configuration of the atoms in *A* or/and *C* sites of Heusler alloy. (b) Atomic arrangement (top panel) and configuration type (bottom panel) of $Ni_{50}Mn_{5+x}Ga_{35-x}Cu_{10}$ at three different Ga contents. And *A,B,C,D* are the four sublattices of Heusler alloy.

FIG. 4. (Color online) Structural and magnetic phase diagram of $Ni_{50}Mn_{5+x}Ga_{35-x}Cu_{10}$ alloys. PM, FM and FIM mean paramagnetic, ferromagnetic and ferrimagnetic states, respectively. $A_p$ and $M_p$ are austenitic /martensitic transformation peak temperature. The ΔT (defined as $M_p$-$A_p$) represents transition temperature hysteresis. $T_C^M$ means the Curie temperature of martensite (all martensitic transition character temperatures in supplementary TABLE SI). And *A,B,C,D* in the insets are the four magnetic sublattices of Heusler alloy.

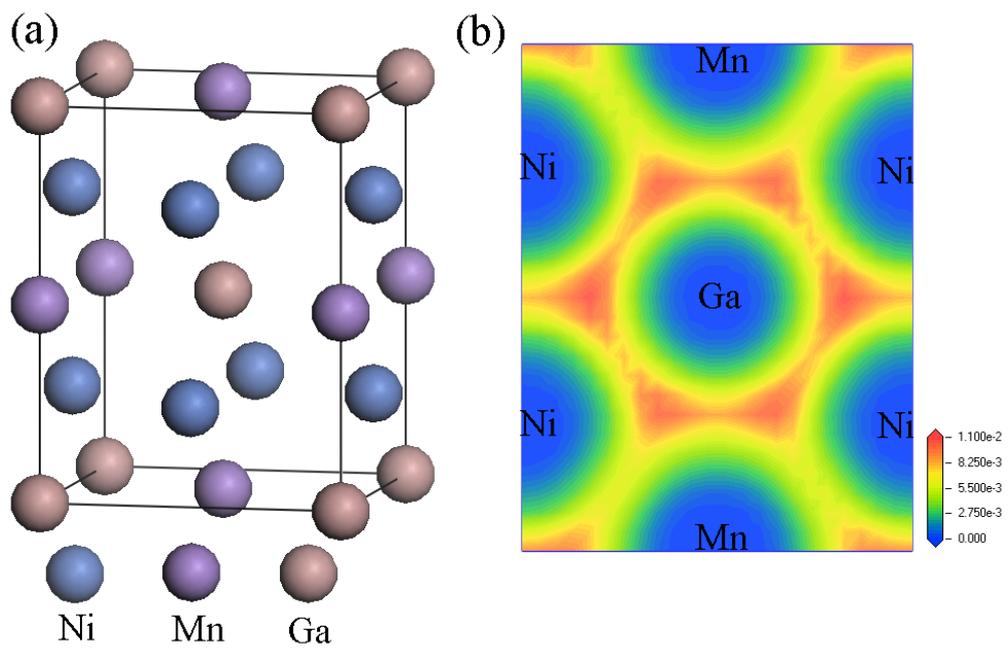

Figure 1, G. J. Li et al., for APL

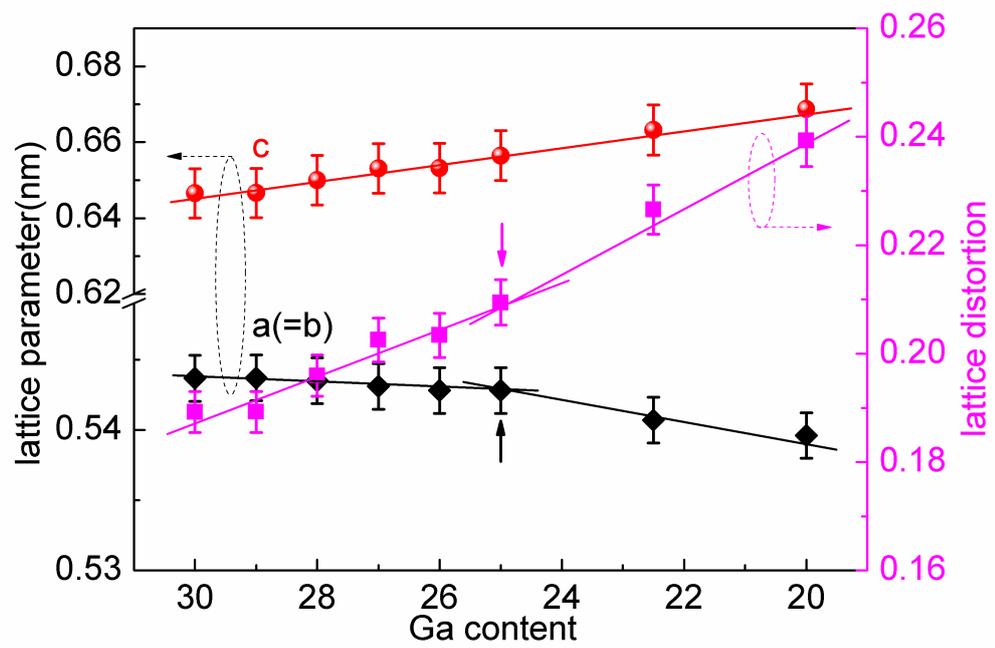

Figure 2 G. J. Li et al., for APL

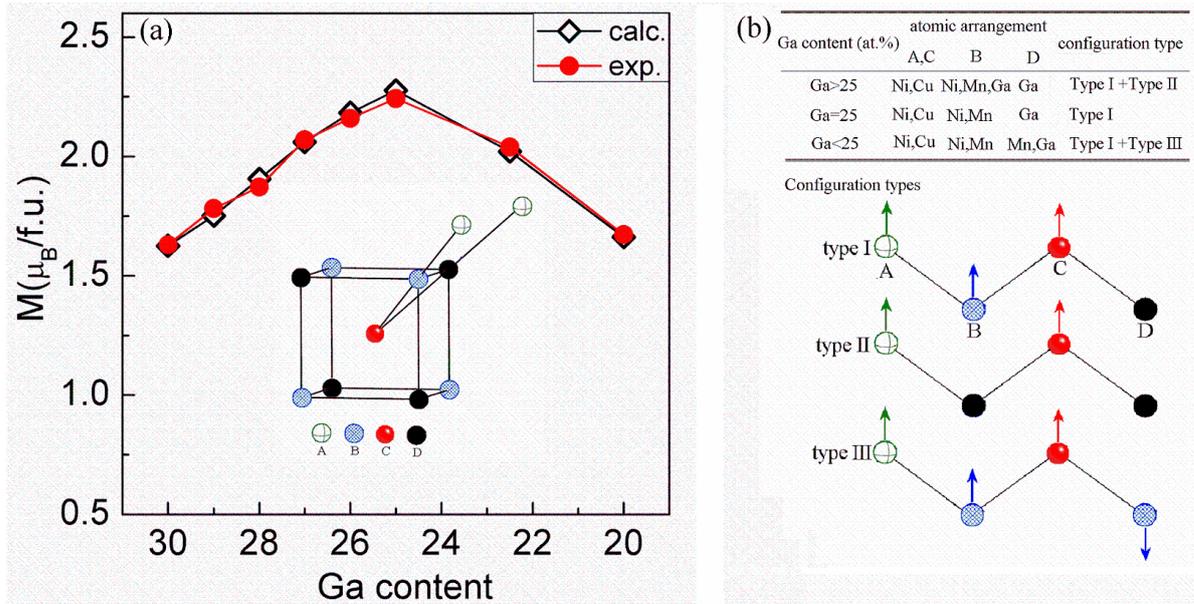

Figure 3 G. J. Li et al., for APL

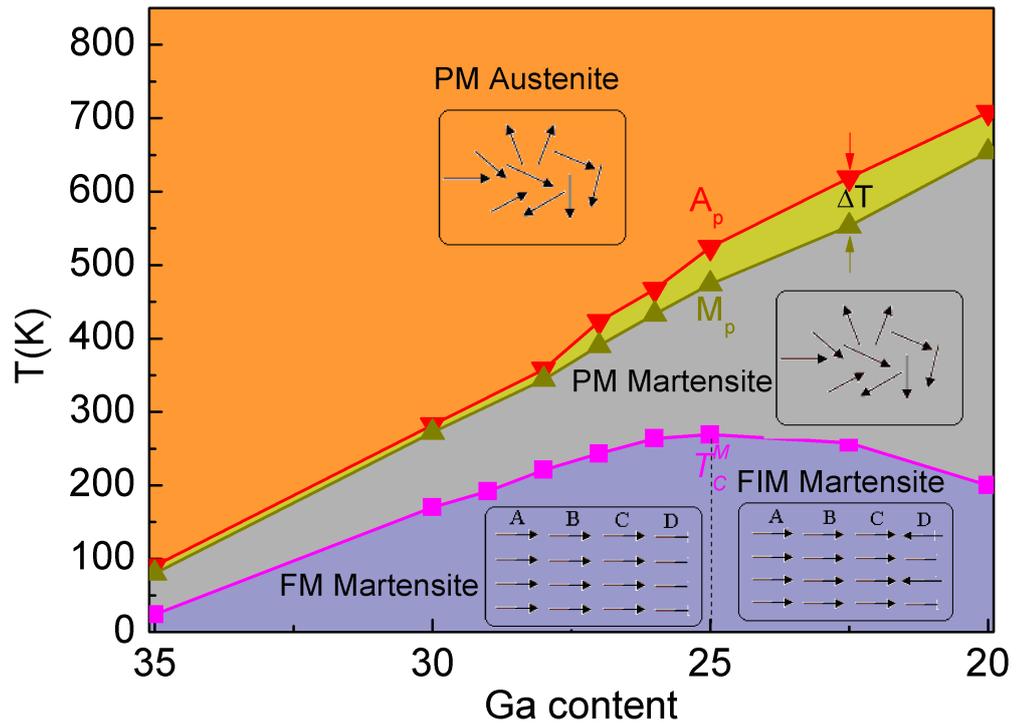

Figure 4, G. J. Li et al., for APL

TABLE I Composition dependence of the calculated atomic (in $\mu_B$/atom) and molecular moments (in $\mu_B$/f.u.) and measured molecular moments (in $\mu_B$/f.u.) for $Ni_{50}Mn_{5-x}Ga_{35-x}Cu_{10}$ alloys.

| Ga (at.%) | Ni(A,C) | Cu(A,C) | Ni(B) | Mn(B) | Mn(D) | Ga(B) | Ga(D) | calc. | exp. |
|---|---|---|---|---|---|---|---|---|---|
| 30 | 0.17 | 0.03 | 0.11 | 3.32 | -- | 0.0003 | -0.03 | 1.62 | 1.63 |
| 27 | 0.21 | 0.04 | 0.12 | 3.24 |  | -0.0012 | -0.04 | 2.06 | 2.07 |
| 25 | 0.25 | 0.08 | 0.09 | 3.04 | -- | -- | -0.04 | 2.28 | 2.24 |
| 22.5 | 0.22 | 0.03 | 0.12 | 3.42 | -3.57 |  | -0.04 | 2.02 | 2.04 |
| 20 | 0.20 | 0.08 | 0.18 | 3.28 | -3.51 |  | -0.04 | 1.66 | 1.67 |

# Supplementary Information

Role of covalent hybridization in martensitic structure and magnetic properties of shape memory alloys: the case of $Ni_{50}Mn_{5+x}Ga_{35-x}Cu_{10}$

G. J. Li,[1] E. K. Liu,[1] H. G. Zhang,[1] Y. J. Zhang,[1] G. Z Xu,[1] H.Z.Luo,[2] H. W. Zhang,[1] W. H. Wang,[2,a)] and G. H. Wu[1]

*1 Beijing National Laboratory for Condensed Matter Physics, Institute of Physics, Chinese Academy of Sciences, Beijing 100190, People's Republic of China*

*2 School of Material Science and Engineering, Hebei University of Technology, Tianjin 300130, People's Republic of China*

a) Author to whom correspondence should be addressed; electronic mail: wenhong.wang@iphy.ac.cn

I. XRD patterns of $Ni_{50}Mn_{5+x}Ga_{35-x}Cu_{10}$ alloys

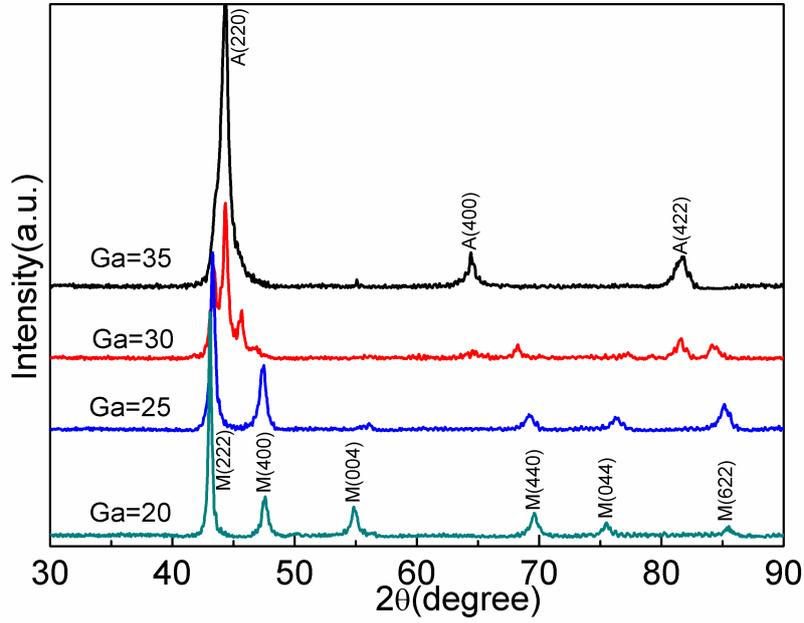

FIG. S1 X-ray diffraction patterns of $Ni_{50}Mn_{5+x}Ga_{35-x}Cu_{10}$ alloys($x$=0-15), in which A and M denote austenite and martensite, respectively.

The XRD patterns of $Ni_{50}Mn_{5+x}Ga_{35-x}Cu_{10}$ alloys measured at room temperature are shown in Fig.S1. One can find that the crystal structure of these alloys changes from B2 austenitic structure (Ga(at.%)≧30) to non-modulated $L1_0$ martensitic structure (Ga(at.%)≦30) with decreasing Ga content. Especially when Ga(at.%) content ranges from 30 to 20, the single $L1_0$ martensite can be maintained in this composition window. Thus, the single $L1_0$ martensitic structure will provide an advantage to carry out the study in this letter.

II. Martensitic transformation character temperatures of $Ni_{50}Mn_{5+x}Ga_{35-x}Cu_{10}$ alloys.

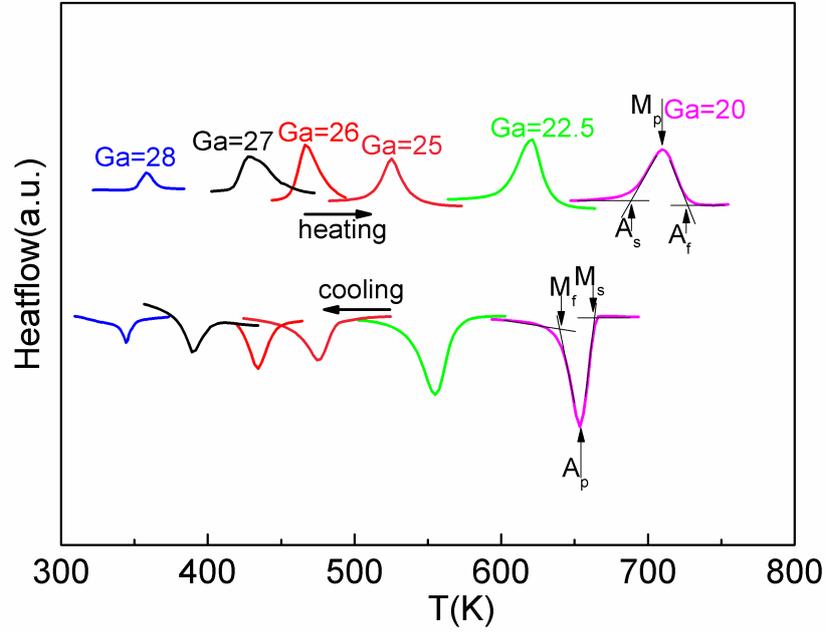

FIG.S2 DSC curves of $Ni_{50}Mn_{5+x}Ga_{35-x}Cu_{10}$ alloys, where $A_s/M_s$, $A_p/M_p$, $A_f/M_f$ mean austenitic /martensitic transformation starting temperature, austenitic /martensitic transformation peak temperature and austenitic /martensitic transformation finishing temperature, respectively.

Figure S2 shows the DSC curves of $Ni_{50}Mn_{5+x}Ga_{35-x}Cu_{10}$ alloys in the vicinity of the martensitic transformation. Actually, all the measurements were carried out ranging the temperature window from 300 K to 1273 K. The austenitic /martensitic transformation starting temperature $A_s/M_s$, austenitic /martensitic transformation peak temperature $A_p/M_p$, and austenitic /martensitic transformation finishing temperature $A_f/M_f$ were determined from the these curves by the method shown in Fig.S2. All the character temperatures were obtained and shown in TABLE.SI.

TABLE SI. Valence electronic concentrations (e/a), martensitic transformation character temperatures, temperature hysteresis ($\Delta T$), transformation latent heat ($\Delta Q$) and martensitic Curie temperature ($T_C^M$) of $Ni_{50}Mn_{5+x}Ga_{35-x}Cu_{10}$ alloys

| Ga(at.%) | e/a | $A_s$ (K) | $A_f$ (K) | $A_p$ (K) | $M_s$ (K) | $M_f$ (K) | $M_p$ (K) | $\Delta T$ (K) | $\Delta Q$ (J/g) | $T_C^M$ (K) |
|---|---|---|---|---|---|---|---|---|---|---|
| 35 | 7.5 | 64 | 116 | 90 | 103 | 57 | 80 | 10 | -- | 24 |
| 30 | 7.7 | 247 | 317 | 282 | 304 | 240 | 272 | 10 | -- | 170 |
| 28 | 7.78 | 350 | 367 | 358 | 349 | 339 | 345 | 13 | 3.01 | 221 |
| 27 | 7.82 | 410 | 443 | 423 | 403 | 380 | 390 | 33 | 3.81 | 243 |
| 26 | 7.86 | 455 | 486 | 467 | 446 | 419 | 433 | 34 | 4.69 | 264 |
| 25 | 7.9 | 513 | 544 | 525 | 490 | 457 | 475 | 50 | 5.97 | 269 |
| 22.5 | 8.0 | 600 | 635 | 620 | 567 | 537 | 555 | 65 | 8.93 | 258 |
| 20 | 8.1 | 690 | 728 | 710 | 665 | 640 | 653 | 57 | 8.62 | 200 |

III. Isothermal magnetization curves of $Ni_{50}Mn_{5+x}Ga_{35-x}Cu_{10}$ alloys

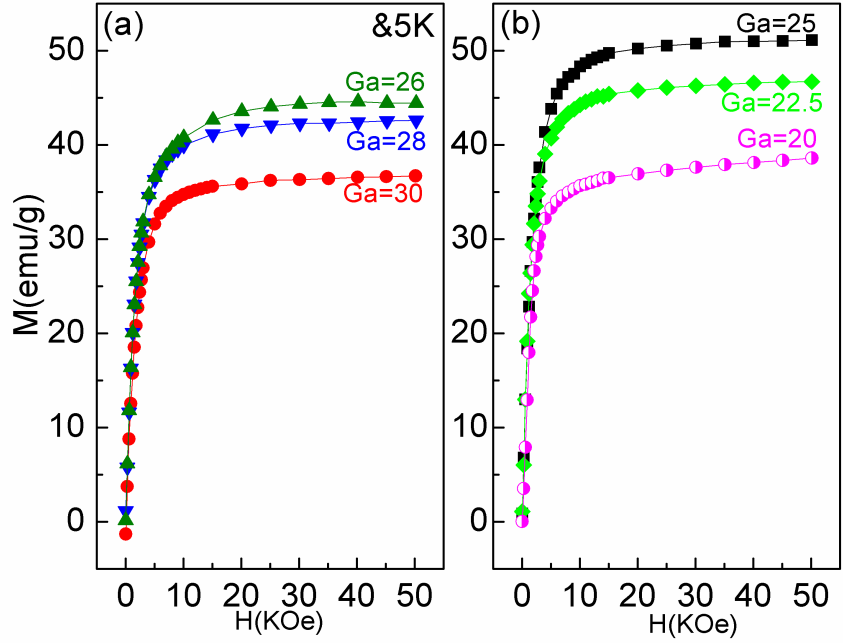

FIG.S3 Isothermal magnetization curves of $Ni_{50}Mn_{5+x}Ga_{35-x}Cu_{10}$ alloys measured at 5K. The left panel (a) shows the isothermal magnetization curves for the alloys with Ga content (at.%) > 25 and right panel (b) shows the isothermal magnetization curves for the alloys with Ga content (at.%) ≤ 25.

The magnetic field dependence of magnetization curves of $Ni_{50}Mn_{5+x}Ga_{35-x}Cu_{10}$ alloys were shown in Fig.S3. One can find that the magnetization value increases with the decrease of Ga content and reaches a maximum at Ga(at.%) =25. And then unusually decrease with the further decrease of Ga content. In this work, the magnetization values under the magnetic field of 50 KOe were chosen as the saturation magnetization values.

IV. Temperature dependence of magnetization curves of $Ni_{50}Mn_{5+x}Ga_{35-x}Cu_{10}$ alloys.

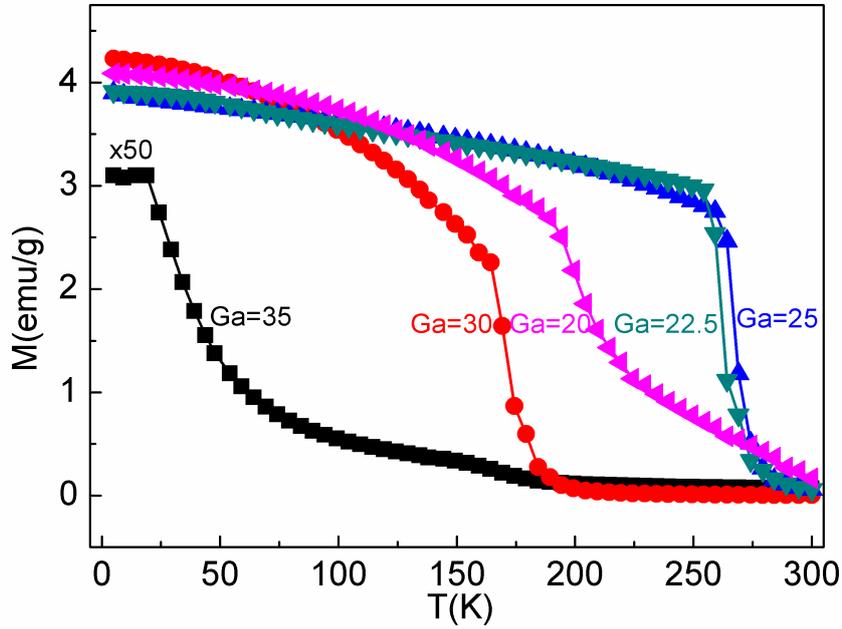

FIG.S4 Temperature dependence of magnetization of $Ni_{50}Mn_{5+x}Ga_{35-x}Cu_{10}$ alloys measured under a magnetic field of 100 Oe.

Fig.S4 shows the temperature dependence of magnetization of $Ni_{50}Mn_{5+x}Ga_{35-x}Cu_{10}$ alloys. In this work, the minimums in the dM(T)/dT curves were chosen as Curie temperatures. From Fig.S4, we can find that the Curie temperature of $Ni_{50}Mn_{5+x}Ga_{35-x}Cu_{10}$ alloys increases with the decrease of Ga content, reaching a maximum at Ga(at.%)=25, and then decreases with the further decrease of Ga content.